\documentclass[twoside]{dis09}
\usepackage[latin1]{inputenc}
\usepackage[dvips]{graphicx,epsfig,color}
\usepackage{wrapfig,rotating}
\usepackage{amssymb,amsmath,array}

\begin{document}
\title{Deeply Virtual Compton Scattering at HERA and 
perspectives at CERN}

\author{Laurent Schoeffel
%
\vspace{.3cm}\\
%
CEA Saclay, Irfu/SPP, 91191 Gif-sur-Yvette Cedex, France
}

\maketitle

\begin{abstract}
Standard parton distribution functions contain neither information on the
correlations between partons nor on their transverse motion,
then a vital knowledge about the three dimensional 
structure of the nucleon is lost.
Hard exclusive processes, in particular DVCS, are essential reactions to go beyond
this standard picture.
In the following, we examine the most recent data from HERA 
(at low $x_{Bj}<10^{-2}$) and their
impact on GPD models. 
The most recent measurements of the Beam Charge Asymmetry 
by the H1 experiment is discussed.
Perspectives are presented 
for further measurements of
DVCS cross sections at CERN, within the COMPASS experiment.
\end{abstract}

\section{Introduction}

Measurements of the deep-inelastic scattering (DIS) of leptons and nucleons, $e+p\to e+X$,
allow the extraction of Parton Distribution Functions (PDFs) which describe
the longitudinal momentum carried by the quarks, anti-quarks and gluons that
make up the fast-moving nucleons. 
These functions have been measured over a wide
kinematic range in the Bjorken scaling variable $x_{Bj}$ and
the photon virtuality $Q^2$.
While PDFs provide crucial input to
perturbative Quantum Chromodynamic (QCD) calculations of processes involving
hadrons, they do not provide a complete picture of the partonic structure of
nucleons. 
In particular, PDFs contain neither information on the
correlations between partons nor on their transverse motion,
then a vital knowledge about the three dimensional 
structure of the nucleon is lost.
Hard exclusive processes, in  which the
nucleon remains intact, have emerged in recent years as prime candidates to complement
this essentially one dimentional picture. 

The simplest exclusive process is the deeply virtual
Compton scattering (DVCS) or exclusive production of real photon, $e + p \rightarrow e + \gamma + p$.
This process is of particular interest as it has both a clear
experimental signature and is calculable in perturbative QCD. 
The DVCS reaction can be regarded as the elastic scattering of the
virtual photon off the proton via a colourless exchange, producing a 
real photon in the final state  \cite{dvcsh1,dvcszeus}. 
In the Bjorken scaling 
regime, 
QCD calculations assume that the exchange involves two partons, having
different longitudinal and transverse momenta, in a colourless
configuration. These unequal momenta or skewing are a consequence of the mass
difference between the incoming virtual photon and the outgoing real
photon. This skewedness effect can
 be interpreted in the context of generalised
parton distributions (GPDs) \cite{disp}. These functions
carry information on both the longitudinal and the
transverse distribution of partons.
The DVCS cross section depends, therefore, on GPDs \cite{disp}.
In the following, we examine the most recent data recorded from the DESY $ep$
collider at HERA and their implication
on models \cite{dvcsh1,dvcszeus}.

\section{Latest experimental measurements from HERA at low $x_{Bj}$}

\begin{figure}[!] 
  \begin{center}
    \includegraphics[width=8.cm]{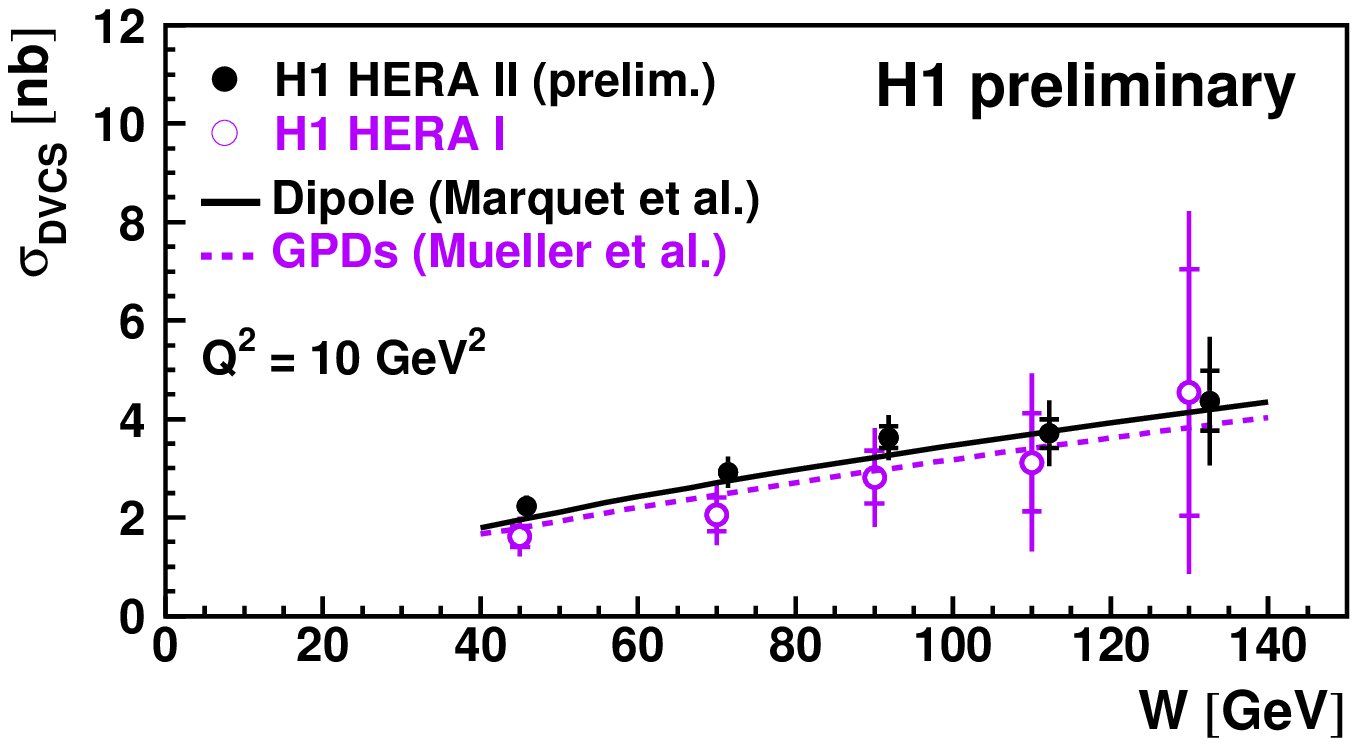}
    \includegraphics[width=8.cm]{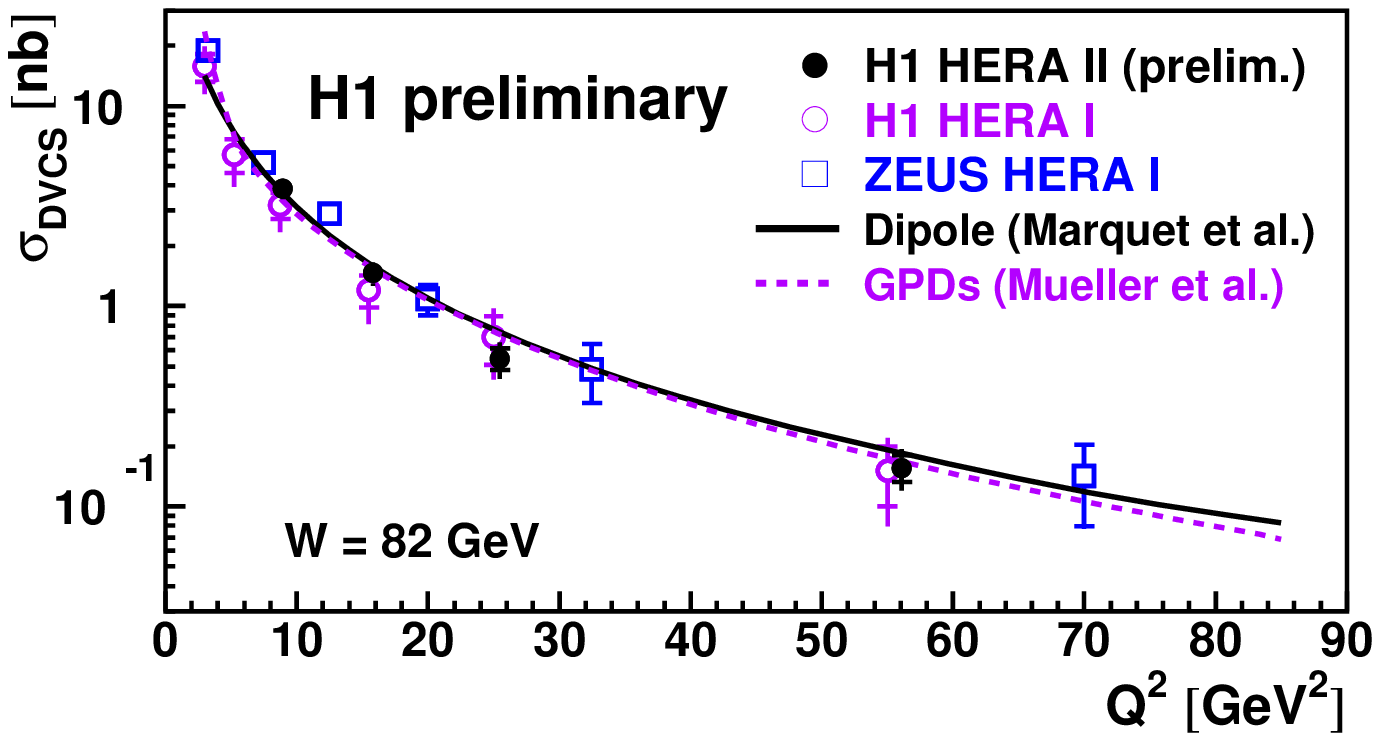}
  \end{center}
  \caption{DVCS cross section for the full HERA data as a function of
$W$ and $Q^2$.
}
\label{fig1}  
\end{figure}

The first measurements of DVCS cross section have been realised  at HERA within 
the H1 and
ZEUS experiments \cite{dvcsh1,dvcszeus}. These results are given in the specific
 kinematic domain
of both experiments,
at low $x_{Bj}$ ($x_{Bj} < 0.01$) but they take advantage of the large range in $Q^2$, 
offered by the
HERA kinematics, which covers more than 2 orders
of magnitude, from $1$ to $100$ GeV$^2$. It makes possible to study the transition from
the low $Q^2$ non-perturbative region (around $1$ GeV$^2$) towards higher values 
of $Q^2$ where the higher twists
effects are lowered (above $10$ GeV$^2$).
The last DVCS cross sections as a functon of
$Q^2$ and $W \simeq \sqrt{Q^2/x}$ are presented 
on Fig. \ref{fig1}. A good agreement with GPDs \cite{disp} 
and dipole \cite{dipole} models is observed.
A very fundamental observation is the steep $W$ dependence in $W^{0.7}$, visible
on Fig. \ref{fig1}. This means that DVCS is a hard process. Thus, 
it is justified to compare DVCS measurements with perturbative QCD calculations,
 GPDs or dipole approaches, as displayed in Fig. \ref{fig1}.

\begin{figure}[!] 
  \begin{center}    
    \includegraphics[width=6cm]{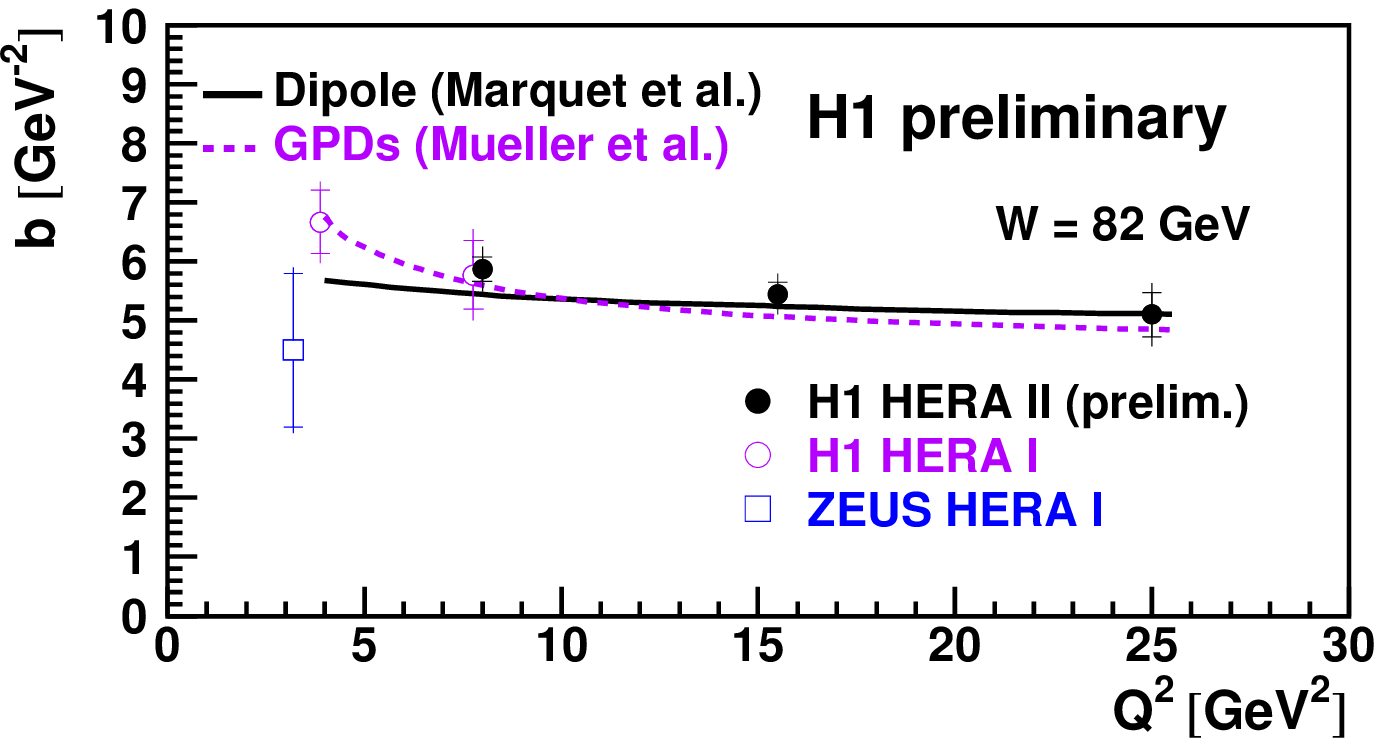}
    \includegraphics[width=6cm]{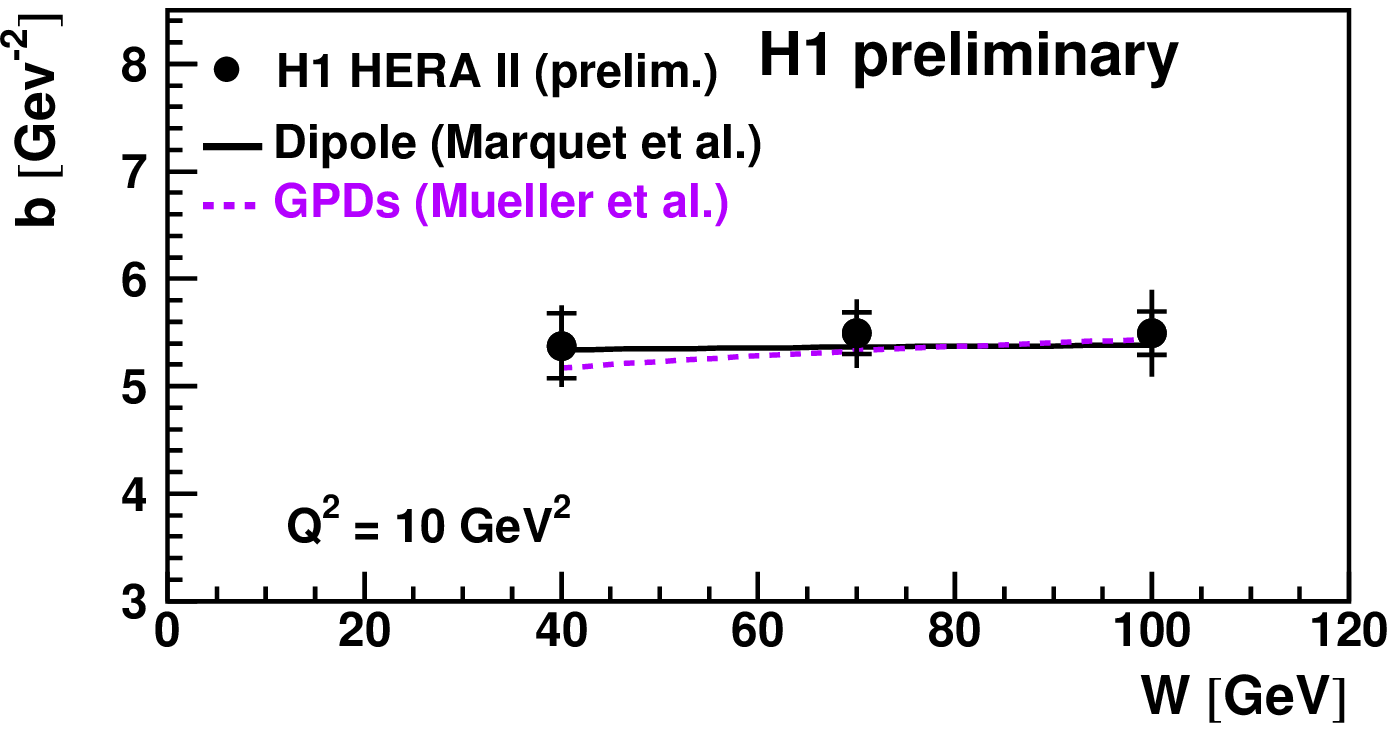}
  \end{center}
  \caption{The logarithmic slope of the $t$ dependence
  for DVCS exclusive production, $b$ as a function of $Q^2$ and $W$, extracted from a fit
  $d\sigma/dt \propto
\exp(-b|t|)$  where $t=(p-p')^2$.
}
\label{fig2}  
\end{figure}

A major experimental achievement 
of H1 and ZEUS \cite{dvcsh1,dvcszeus} has been the measurement of
DVCS cross sections, differential in $t=(p'-p)^2$, 
the momentum transfer (squared) at the proton vertex.
A good description
of $d\sigma_{DVCS}/dt$ by a fit of the form $e^{-b|t|}$
is obtained \cite{dvcsh1,dvcszeus}. 
Hence, an extraction of the $t$-slope parameter $b$ is accessible
and it can be achieved experimentally
for different values of $Q^2$ and $W$ (see Fig. \ref{fig2}).
Again, we observe the good agreement 
of measurements with GPDs and dipole models.

\section{Nucleon Tomography and Perspectives at CERN}

Measurements of the $t$-slope parameters $b$
are key measurements for almost all exclusive processes,
in particular DVCS. Indeed,
a Fourier transform from momentum
to impact parameter space readily shows that the $t$-slope $b$ is related to the
typical transverse distance between the colliding objects \cite{disp}.
At high scale, the $q\bar{q}$ dipole is almost
point-like, and the $t$ dependence of the cross section is given by the transverse extension 
of the gluons (or sea quarks) in the  proton for a given $x_{Bj}$ range.
More precisely, from GPDs, we can compute
a parton density which also depends on a spatial degree of freedom, the transverse size (or impact parameter), labeled $R_\perp$,
in the proton. Both functions are related by a Fourier transform 
$$
PDF (x, R_\perp; Q^2) 
\;\; \equiv \;\; \int \frac{d^2 \Delta_\perp}{(2 \pi)^2}
\; e^{i ({\Delta}_\perp {R_\perp})}
\; GPD (x, t = -{\Delta}_\perp^2; Q^2).
$$
Thus, the transverse extension $\langle r_T^2 \rangle$
 of gluons (or sea quarks) in the proton can be written as
$$
\langle r_T^2 \rangle
\;\; \equiv \;\; \frac{\int d^2 R_\perp \; PDF(x, R_\perp) \; R_\perp^2}
{\int d^2 R_\perp \; PDF(x, R_\perp)} 
\;\; = \;\; 4 \; \frac{\partial}{\partial t}
\left[ \frac{GPD (x, t)}{GPD (x, 0)} \right]_{t = 0} = 2 b
$$
where $b$ is the exponential $t$-slope.
Measurements of  $b$
presented in Fig. \ref{fig2}
corresponds to $\sqrt{r_T^2} = 0.65 \pm 0.02$~fm at large scale 
$Q^2$ for $x_{Bj} < 10^{-2}$.
This value is smaller that the size of a single proton, and, 
in contrast to hadron-hadron scattering, it does not expand as energy $W$ increases.
This result is consistent with perturbative QCD calculations in terms 
of a radiation cloud of gluons and quarks
emitted around the incoming virtual photon.
The fact the perturbative QCD calculations provide correct descriptions
of $b$ measurements (see previous section) is a proof that
they deal correctly this this non-trivial aspect of the proton 
(spatial) structure. The correlation between the spatial transverse
structure and the longitudinal momenta distributions of partons in the proton
is one major challenge of the GPDs model. 

\begin{figure}[htbp] 
  \begin{center}
    \includegraphics[width=10cm]{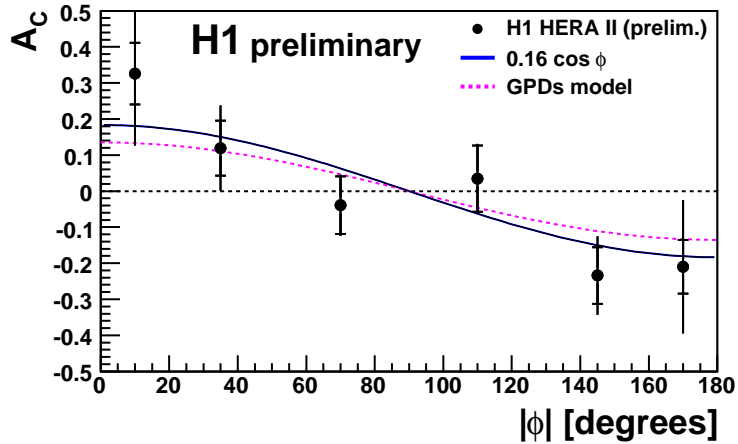}
  \end{center}
  \caption{Beam charge asymmetry as a function of $\phi$ measured by H1.
  Statistical and systematical uncertainties are shown. Data are corrected 
  from the migrations of events in $\phi$.
  A comparison with the GPDs model described in Ref. \cite{disp} is presented.
  It fits very nicely with the best fit to the data, in $p_1 \cos \phi$
  ($p_1=0.16$).
}
\label{fig3}  
\end{figure}

Another natural experimental way to address this problem is 
proceeds from
a determination of a cross section asymmetry with respect to the beam
charge. It has been realised recently by the H1 experiment by measuring the ratio
$(d\sigma^+ -d\sigma^-)/ (d\sigma^+ + d\sigma^-)$ as a function of $\phi$,
where $\phi$ is the azimuthal angle between leptons and proton plane.
The result is presented on Fig. \ref{fig3} with  a fit in $\cos \phi$.
After applying a deconvolution method to account for the  resolution on $\phi$,
the coefficient of the $\cos \phi$ dependence is found to 
be $p_1 = 0.16 \pm 0.03 (stat.) \pm 0.05 (sys.)$ (at low $x_{Bj}<0.01$).
This result represents obviously a major experimental progress.
Using present HERA data at low $x_{Bj}$,
as well as JLab and HERMES data at larger $x_{Bj}$ ($x_{Bj}>0.1$), 
a first global parametrisation of GPDs can be done \cite{disp}.
This is an essential step forward in the field.
However, some efforts have
still to be made in the intermediate $x_{Bj}$ domain.

Feasabilities for future Beam Charge Asymmetry (BCA) 
measurements at COMPASS have been studied extensively
in the last decade \cite{dhose}.
 COMPASS is a fixed target experiment which can use
100 GeV muon beams and hydrogen targets, and then access 
experimentaly the DVCS process $\mu p \rightarrow \mu \gamma p$.
The BCA can be determined when using positive and negative muon beams.
One major interest is the kinematic coverage from $2$ GeV$^2$ till $6$ GeV$^2$ in $Q^2$
and  $x_{Bj}$ ranging from $0.05$ till $0.1$. It means that it is possible to avoid
the kinematic domain dominated by higher-twists and non-perturbative effects 
(for $Q^2 < 1$ GeV$^2$) and keeping a
$x_{Bj}$ range which is extending the HERA (H1/ZEUS) domain.

\begin{figure}[htbp]
\begin{center}
 \includegraphics[totalheight=6cm]{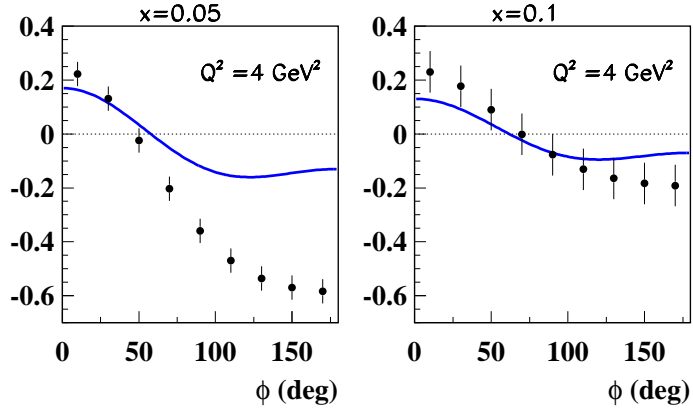}
\end{center}
\caption{\label{lolo}
Simulation of the azimuthal angular distribution of the beam charge
asymmetry measurable at COMPASS at $E_\mu=100$ GeV.
We present the projected values and error bars in the range
$|t|<$ 0.6 GeV$^2$
for 2 values of $x_{Bj}$ ($0.05$ and $0.1$) at $Q^2=4$ GeV$^2$ 
(\cite{dhose}).
The prediction of the GPD model with a non-factorised $t$ dependence 
is shown (full line). The case of a factorised $t$ dependence would lead to
a prediction of the BCA compatible with zero and is not displayed.
 }
\end{figure}

In Fig. \ref{lolo}, we compare QCD predictions 
of the GPDs model used in Ref. \cite{dvcsh1} 
 to simulations of 
the BCA extraction at COMPASS using a muon beam
of 100 GeV \cite{compasslolo}. 
We present the data/theory comparisons for one value of 
$Q^2$ ($4$ GeV$^2$) and two values of
$x_{Bj}$ ($0.05$ and $0.1$). Simulations have been done unsing 
the VGG model to derive BCA values \cite{dhose}. 
As mentioned above, this is obviously an essential measurement,
to be done  in 2011/2012, in order to
cover the full kinematic range. It would give some 
results in the intermediate $x_{Bj}$ range between H1/ZEUS  and 
JLab/HERMES experiments and would allow obviously a further step forward
in GPDs models.

\section{Summary and outlook}
DVCS measurements in the HERA kinematics at low $x_{Bj}$ ($x_{Bj}<0.01$)
are well described by recent GPDs models, which also describe
correctly measurements at larger values of $x_{Bj}$ in
the JLab kinematics.
DVCS measurements in the HERA kinematics 
are also nicely described
within a dipole approach,
which encodes
the non-forward kinematics for DVCS only 
through the different weights coming from the
photon wavefunctions. Recently, H1 and ZEUS 
experiments have also shown that proton 
tomography at low $x_{Bj}$
enters into the experimental domain of high energy physics, with a first 
experimental evidence
that gluons are located at the periphery of the proton. A new frontier in 
understanding
this structure would be possible at CERN within the 
COMPASS experimental setup. Major advances have already been done
on the design of the project and simulation outputs.



\begin{footnotesize}



%

\end{footnotesize}


\end{document}